\documentclass[letter]{jpsj3}
\usepackage{txfonts}
\usepackage{color}
\usepackage{bm}
\title{Deep Learning the Quantum Phase Transitions in Random Two-Dimensional Electron Systems}

\author{Tomoki Ohtsuki$^1$\thanks{ootsuki\_t@msi.co.jp} and Tomi Ohtsuki$^2$\thanks{ohtsuki@sophia.ac.jp}}
\inst{
$^1$ NTT DATA Mathematical Systems Inc, Shinjuku-ku, Tokyo 160-0016, Japan\\
$^2$Physics Division, Sophia University, Chiyoda-ku, Tokyo 102-8554, Japan} 

\abst{
Random electron systems show rich phases such as Anderson insulator,
diffusive metal, quantum Hall and quantum anomalous Hall insulators,
Weyl semimetal, as well as strong/weak topological insulators.
Eigenfunctions of each matter phase have specific features, but
owing to the random nature of systems, determining the matter phase from
eigenfunctions is difficult.  Here, we propose the deep
learning algorithm to capture the features of eigenfunctions.
Localization-delocalization transition, as well as disordered Chern insulator-Anderson
insulator transition, is discussed.
}

\begin{document}
\maketitle

{\it Introduction}--
More than half a century has passed since the discovery of Anderson localization\cite{Anderson58},
and the random electron systems continue to attract theoretical as well as
 experimental interest.
Symmetry classification of topological insulators\cite{Schnyder08,Kitaev09,Hasan10,Qi11}
based on the universality classes of random noninteracting electron systems\cite{Zirnbauer96,altland97}
gives rise to a fundamental question: can we distinguish the random topological insulator from Anderson insulators?
Note that topological numbers are usually defined in the randomness free systems via the
integration of the Berry curvature of Bloch function over the Brillouin zone, although
topological numbers in random systems have recently been proposed.\cite{Sbierski14a,Katsura16,Katsura16a}

\begin{figure}[htbp]
  \begin{center}
    \begin{tabular}{c}
    
     \begin{minipage}{0.23\hsize}
  \begin{center}
   \includegraphics[width=28mm,angle=270]{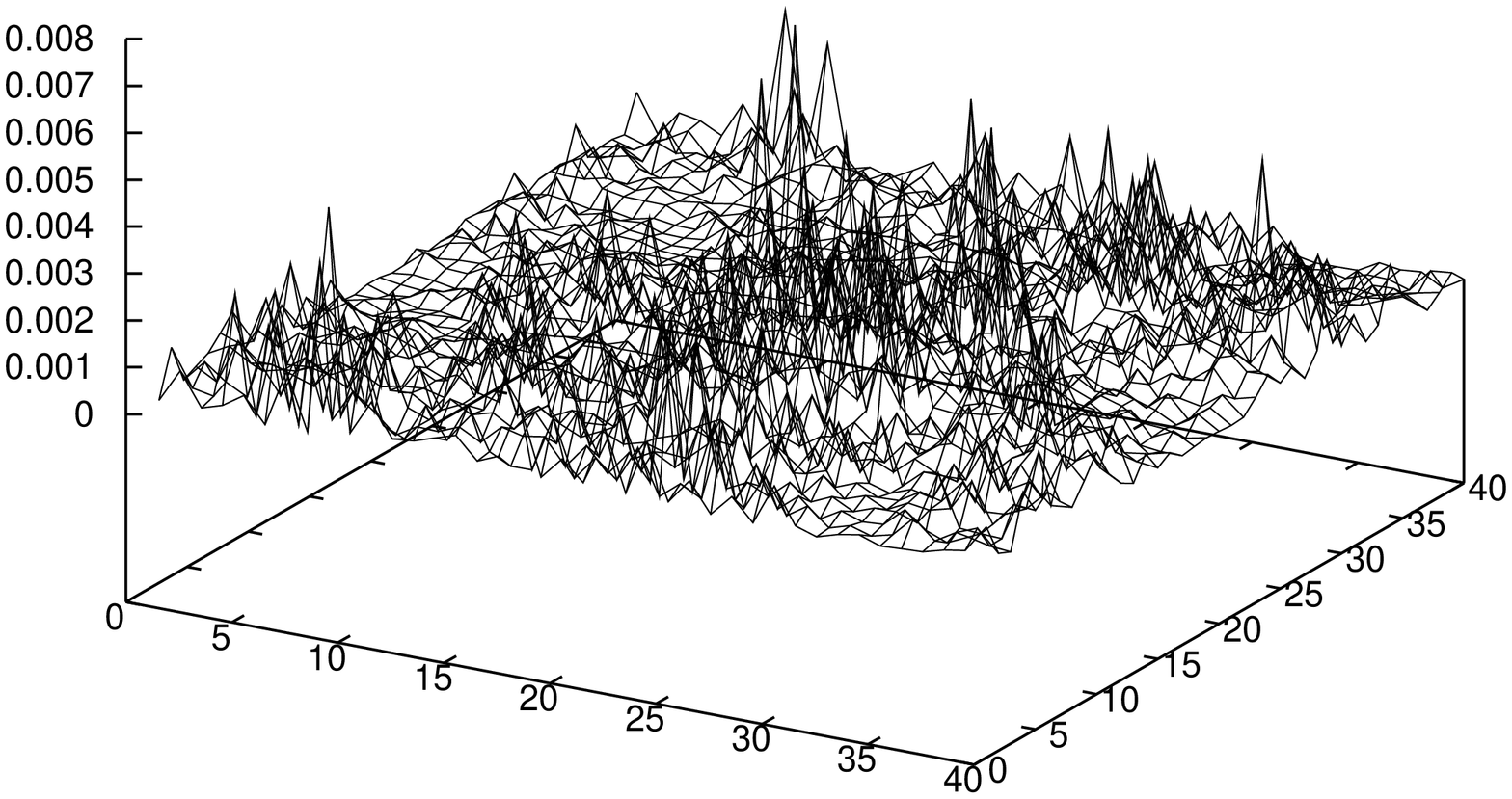}
   \hspace{1.6cm} (a)
     \end{center}
 \end{minipage}
 
 \begin{minipage}{0.23\hsize}
  \begin{center}
   \includegraphics[width=28mm,angle=270]{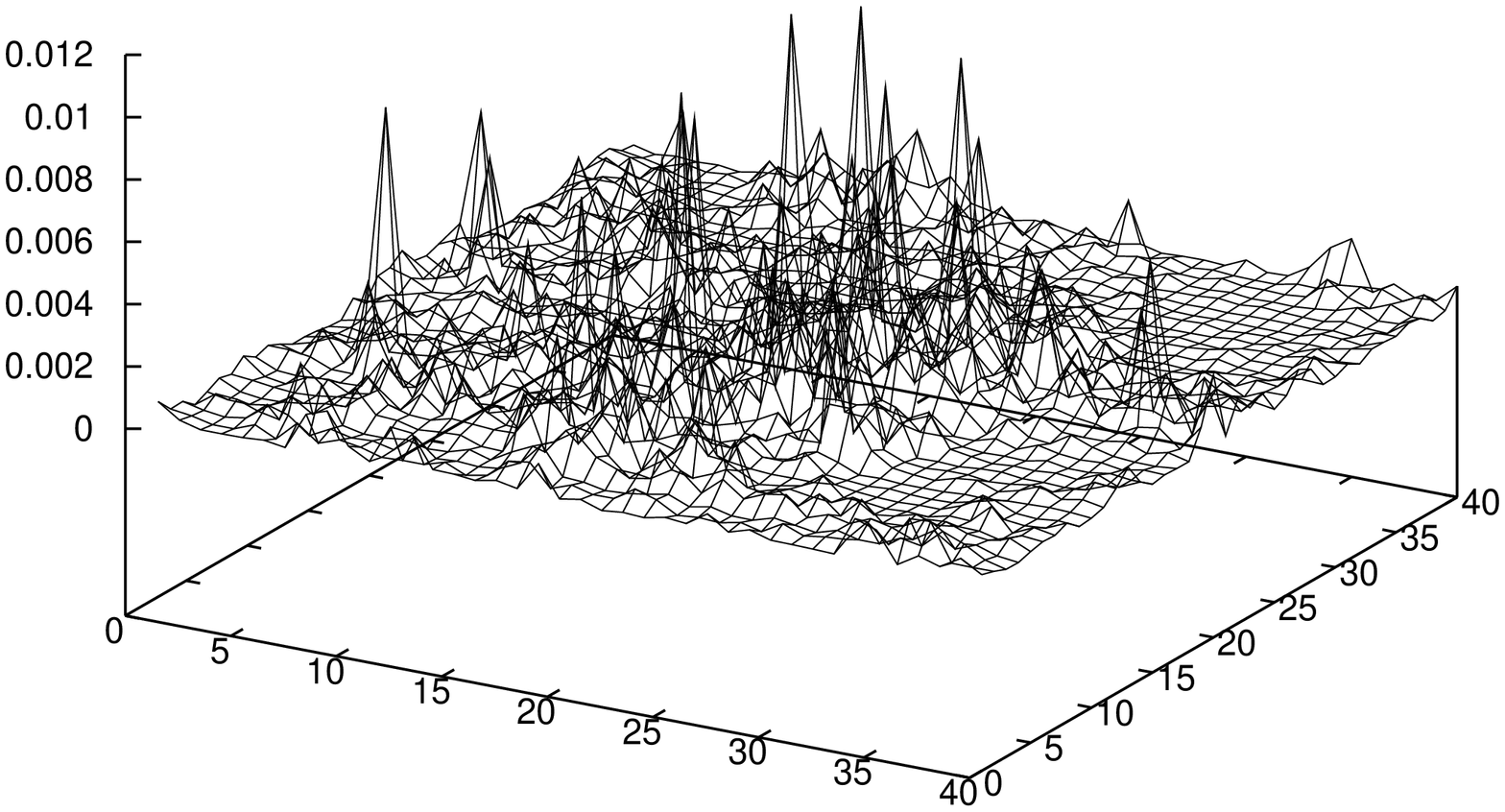}
      \hspace{1.6cm} (b)
     \end{center}
 \end{minipage}

 \begin{minipage}{0.23\hsize}
  \begin{center}
   \includegraphics[width=28mm,angle=270]{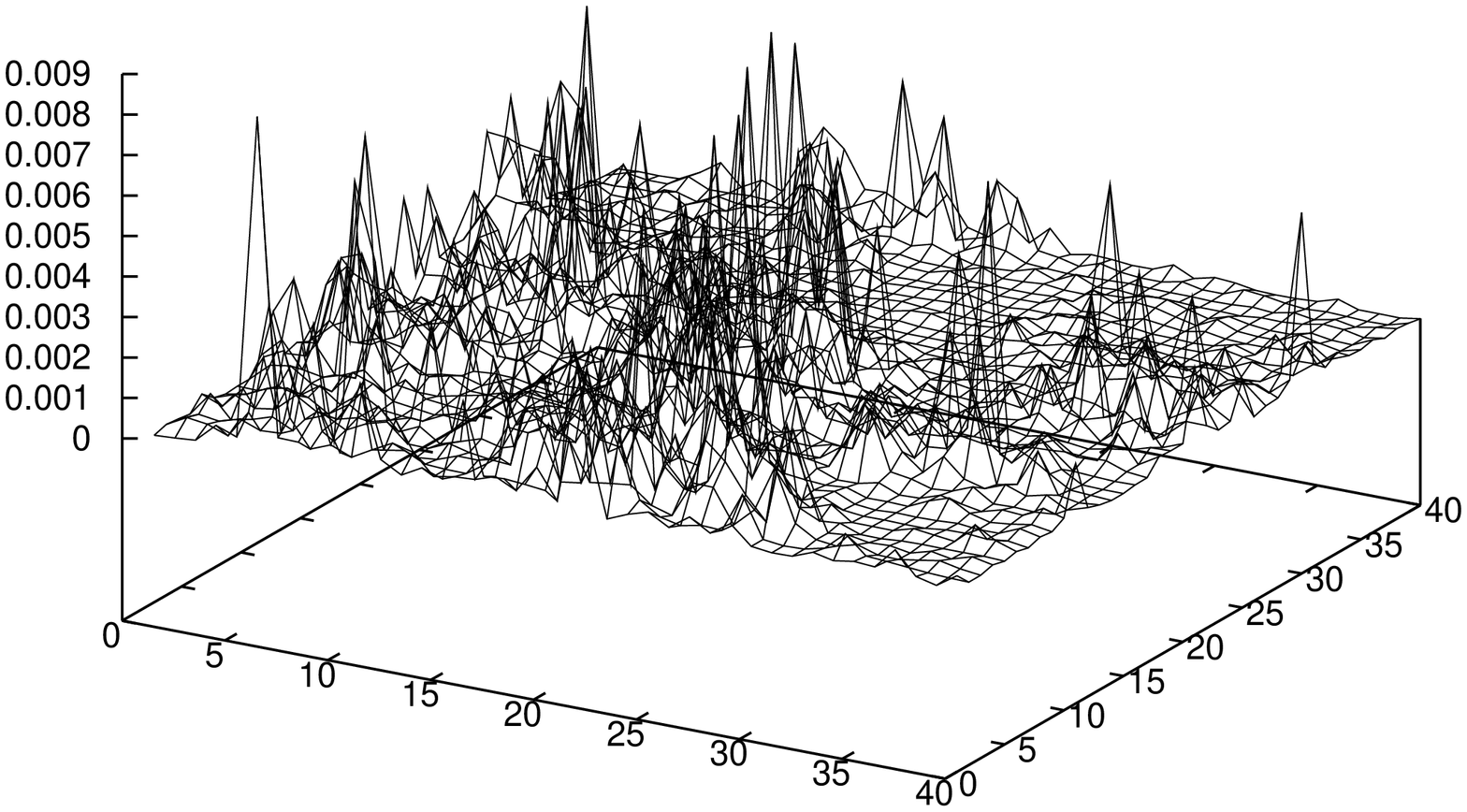}
   \hspace{1.6cm} (c)
  \end{center}
 \end{minipage}

 \begin{minipage}{0.23\hsize}
  \begin{center}
   \includegraphics[width=28mm,angle=270]{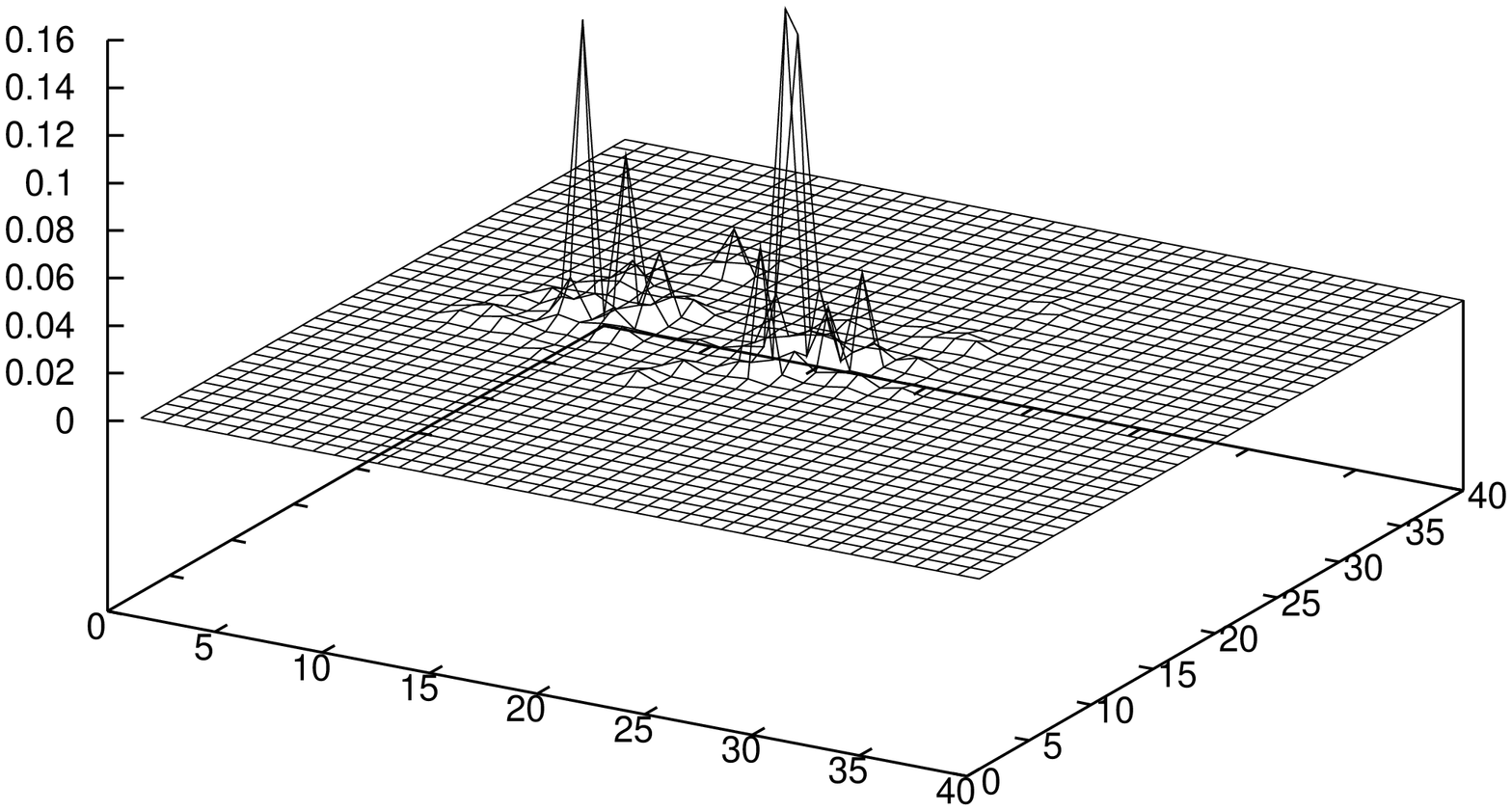}
   \hspace{1.6cm} (d)
  \end{center}
 \end{minipage}
 
  \end{tabular}
 \caption{Examples of eigenfunction modulus squared $|\psi(x,y)|^2$ for
(a) $W=0.124\approx W_c^{\rm SU2}/50$, (b) $W=5.58\approx 0.9W_c^{\rm SU2}$, (c) $W=6.82\approx 1.1 W_c^{\rm SU2}$, and (d) $W=12.4\approx 2W_c^{\rm SU2}$.
Peak positions are shifted to the center of the systems.}
\label{fig:su2Eigenfunctions}
\end{center}
\end{figure}

Determining the phase diagram and the critical exponents requires
large-scale numerical simulation combined with detailed finite size scaling
analyses\cite{alberto10,alberto11,Slevin14,Ujfalusi15}.
This is because, owing to large fluctuations of wavefunction amplitudes,
it is almost impossible to judge whether the eigenfunction obtained by
diagonalizing small systems is localized or delocalized, or
whether the eigenfunction is a chiral/helical edge state of a topological insulator.
In fact, it often happens that eigenfunctions in the localized phase seem
less localized than those in the delocalized phase
[see Figs.~\ref{fig:su2Eigenfunctions}(b) and ~\ref{fig:su2Eigenfunctions}(c) for example].

Recently, there has been  great progress on  image recognition
algorithms\cite{Obuchi14} based on deep machine learning.\cite{LeCun15,Silver16}
Machine learning has recently been applied to several problems of condensed matter physics such as
Ising and spin ice models\cite{Carrasquilla16,Tanaka16}
and strongly correlated systems.\cite{Carleo16,Broecker16,Chng16,Li16,Nieuwenburg16,Huang16}

In this Letter, we test the image recognition algorithm to determine whether the
eigenfunctions for relatively small systems are localized/delocalized, and
topological/nontopological.
As examples, we test two types of two-dimensional (2D) quantum phase transitions:
Anderson-type localization-delocalization transition in symplectic systems,
and disordered Chern insulator to Anderson insulator transition in unitary systems.

{\it Distinguishing Localized States from Delocalized Ones}--
We start with a 2D symplectic system, which is realized in the presence of
spin-orbit scattering.
We use the SU(2) Hamiltonian \cite{Asada02} that describes the
2D electron on a square lattice with nearest-neighbor hopping,
\begin{equation}
\label{eq:su2Hamiltonian}
H=\sum_{i,\sigma}\epsilon_i c_{i,\sigma}^\dagger c_{i,\sigma}-
\sum_{\langle i,j\rangle,\sigma,\sigma'}R(i,j)_{\sigma,\sigma'}
c_{i,\sigma}^\dagger c_{j,\sigma'}\,,
\end{equation}
where $c_{i,\sigma}^\dagger$ ($c_{i,\sigma}$) denotes the creation (annihilation)
operator of an electron at site $i=(x,y)$ with spin $\sigma$, and $\epsilon_i$
denotes the random potential at site $i$. We assume a box distribution with
each  $\epsilon_i$ uniformly and independently distributed on the interval
[$-W/2,W/2]$.
The modulus of the transfer energy is taken to be the energy unit.
$R(i,j)$ is an SU(2) matrix,
\begin{equation}
\label{eq:su2matrix}
R(i, j)=\left(
\begin{array}{cc}
e^{i\alpha_{i,j}}\cos\beta_{i,j}      &   e^{i\gamma_{i,j}}\sin\beta_{i,j}   \\
-e^{-i\gamma_{i,j}}\sin\beta_{i,j}     &   e^{-i\alpha_{i,j}}\cos\beta_{i,j}
\end{array}
\right)\,,
\end{equation}
with $\alpha$ and $\gamma$  uniformly distributed in the range $[0,2\pi)$.
The probability density $P(\beta)$ is 
\begin{equation}
\label{eq:beta}
P(\beta)=\left\{
\begin{array}{ll}
 \sin(2\beta)     &  0\le \beta\le \pi/2  \,,\\
   0   &   {\rm otherwise}\,.
\end{array}
\right.
\end{equation}
Examples of the eigenfunctions in delocalized [Figs.~\ref{fig:su2Eigenfunctions}(a) and \ref{fig:su2Eigenfunctions}(b)] and localized phases [Figs.~\ref{fig:su2Eigenfunctions}(c) and \ref{fig:su2Eigenfunctions}(d)]
are shown in Fig.~\ref{fig:su2Eigenfunctions}.

For $E=0$ (band center), from the finite size scaling analyses of the quasi-1D localization length\cite{Asada02,Asada04}, it is known that
the states are delocalized when $W<W_c^{\rm SU2}(\approx 6.20)$,
while they are localized when $W>W_c^{\rm SU2}$.
We impose periodic boundary conditions in $x$- and $y$-directions,
and diagonalize systems of $40\times 40$.
From the resulting 3200 eigenstates with Kramers degeneracy,
we pick up the 1600th eigenstate (i.e., a state close to the band center).
For simplicity, the maximum modulus of the eigenfunction is shifted to the center of the system.
Changing $W$ and the seed of the random number stream (Intel MKL MT2023),
we prepare 2000 samples of states, i.e., 1000 for $W<W_c^{\rm SU2}$ and 1000 for $W>W_c^{\rm SU2}$.
We then teach the machine whether the states belong to the localized (delocalized) phase.

\begin{figure}[htbp]
    \begin{center}
        \begin{tabular}{cc} 
                (a)
            \begin{minipage}{0.45\hsize} 
                \begin{center} 
                    \includegraphics[width=50mm]{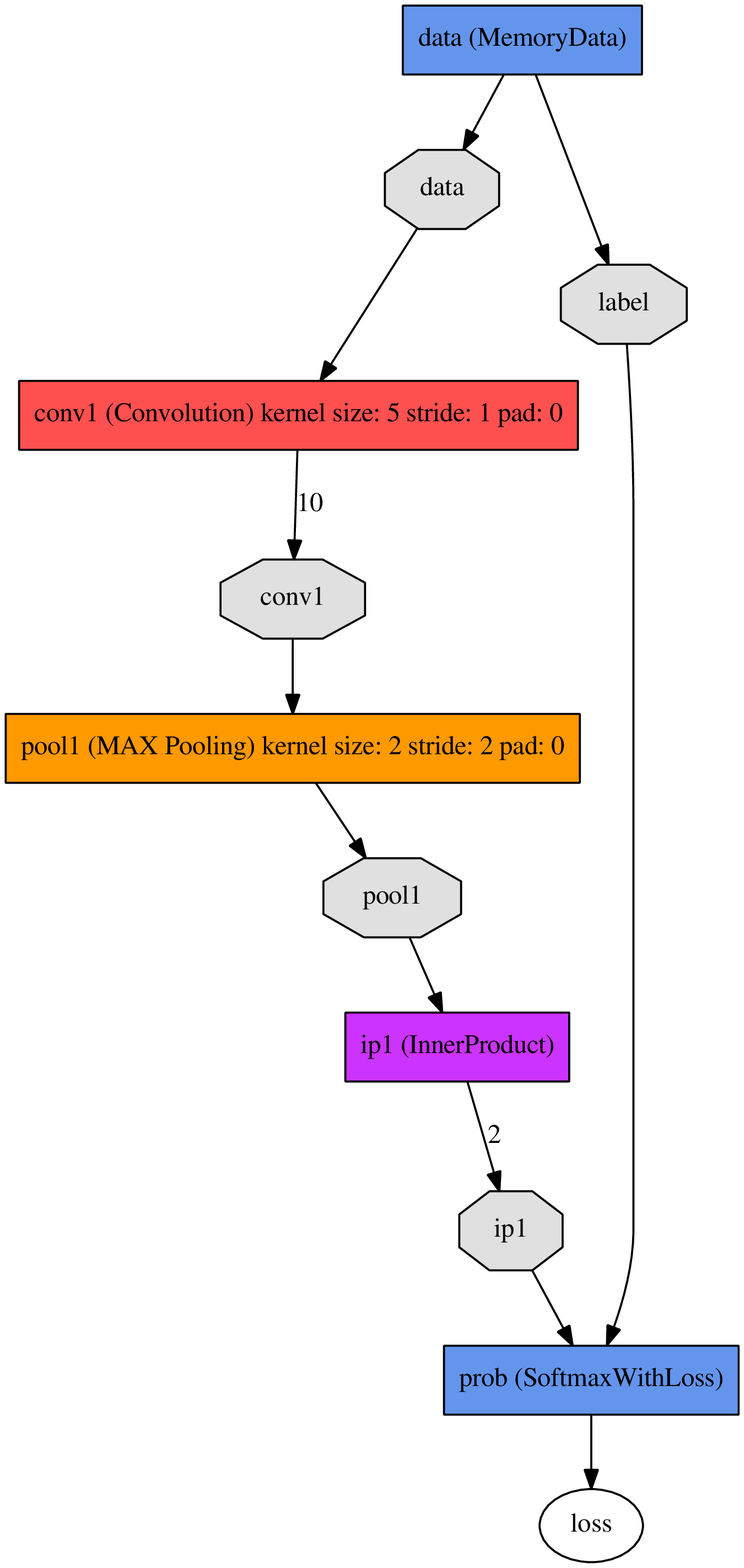}
                \end{center}
            \end{minipage} & 
                    (b) 
            \begin{minipage} {0.45\hsize}
                \begin{center}
                    \includegraphics[width=50mm]{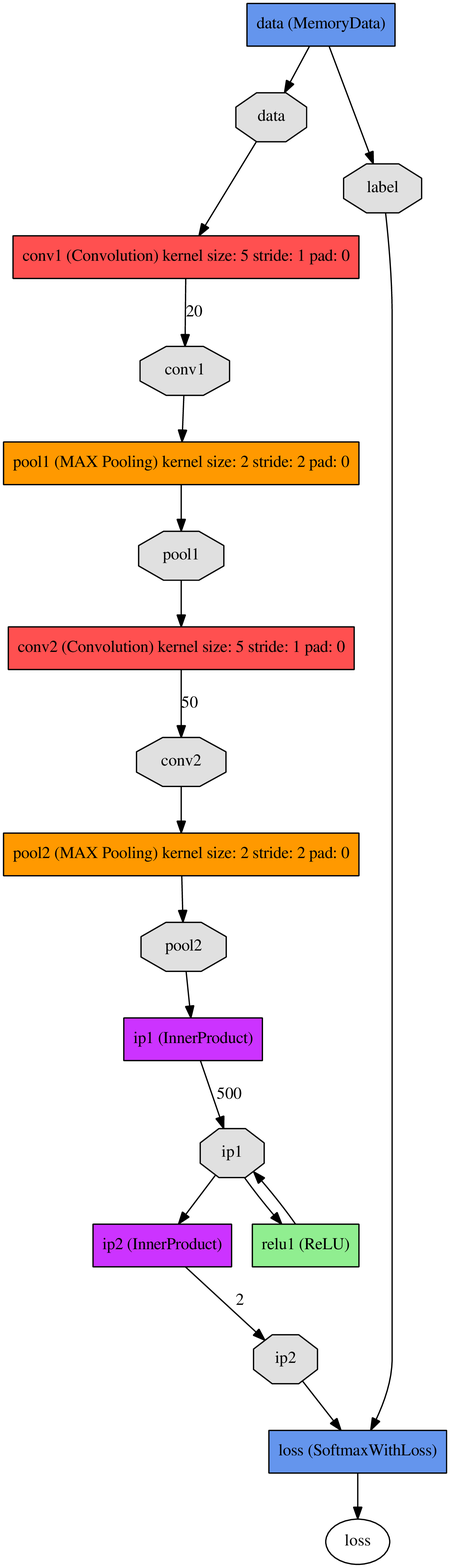}
                \end{center}
            \end{minipage} 
        \end{tabular} 
        \caption{(Color online) Network architectures used in this work. (a) A simple two-weight-layer CNN,
         which consists of convolution and max pooling, followed by dense linear transformation. (b) LeNet-like architecture with ReLU activation. } 
    \label{fig:networkArchitectures}
    \end{center}
\end{figure}

For our network architecture, we consider two types of simple convolutional neural network (CNN), which output two real numbers, i.e., probabilities for each phase, given $40\times 40$ input eigenfunction. The first one is a very simple network with two weight layers, which first convolves the input with a $5\times 5$ filter with stride $1$ to 10 channels, then applies max pooling with a kernel size of $2\times 2$ and stride 2, and finally performs fully connected linear transformation to output the learned probabilities. The loss function can then be defined by the cross entropy of probabilities and the localized/delocalized labels. The second, rather deep one with four weight layers is a variant of LeNet \cite{lecun1998gradient} included in Caffe \cite{jia2014caffe} (with the input size changed to $40\times 40$), which utilizes rectified linear unit (ReLU) as its activation function. See Fig.~\ref{fig:networkArchitectures} for illustration and detailed parameters. 
    The network weight parameters (to be trained) are sampled from gaussian distribution, the scale of which is determined by the number of input and output dimensions\cite{glorot2010understanding}, except for the first convolution layer connected to the raw input: since we are dealing with eigenfunctions, whose typical values at each lattice site are much smaller than those of gray-scale images, we have manually chosen the weight initialization scale to be 100, which worked better in practice for the two networks.
As the stochastic gradient descent solver, we have used the RMSProp solver\cite{tieleman2012lecture} with the parameters in
the Caffe MNIST example (which is contained as \verb+examples/mnist/lenet_solver_rmsprop.prototxt+ in the Caffe source). Before the training, we always partition the training data into 90\% and 10\%, and use the latter as the validation set during the training. The solver performs enough iterations so that the validation error becomes stationary.
We have used a workstation:
Intel Xeon  E5-1620 v4, single CPU with 4 cores with GPU Quadro K420 and
GPGPU TESLA K40 running on Linux CentOS 6.8.

\begin{figure}
  \begin{center}
    \begin{tabular}{c}
         \begin{minipage}{0.45\hsize}
  \begin{center}
   \includegraphics[width=70mm]{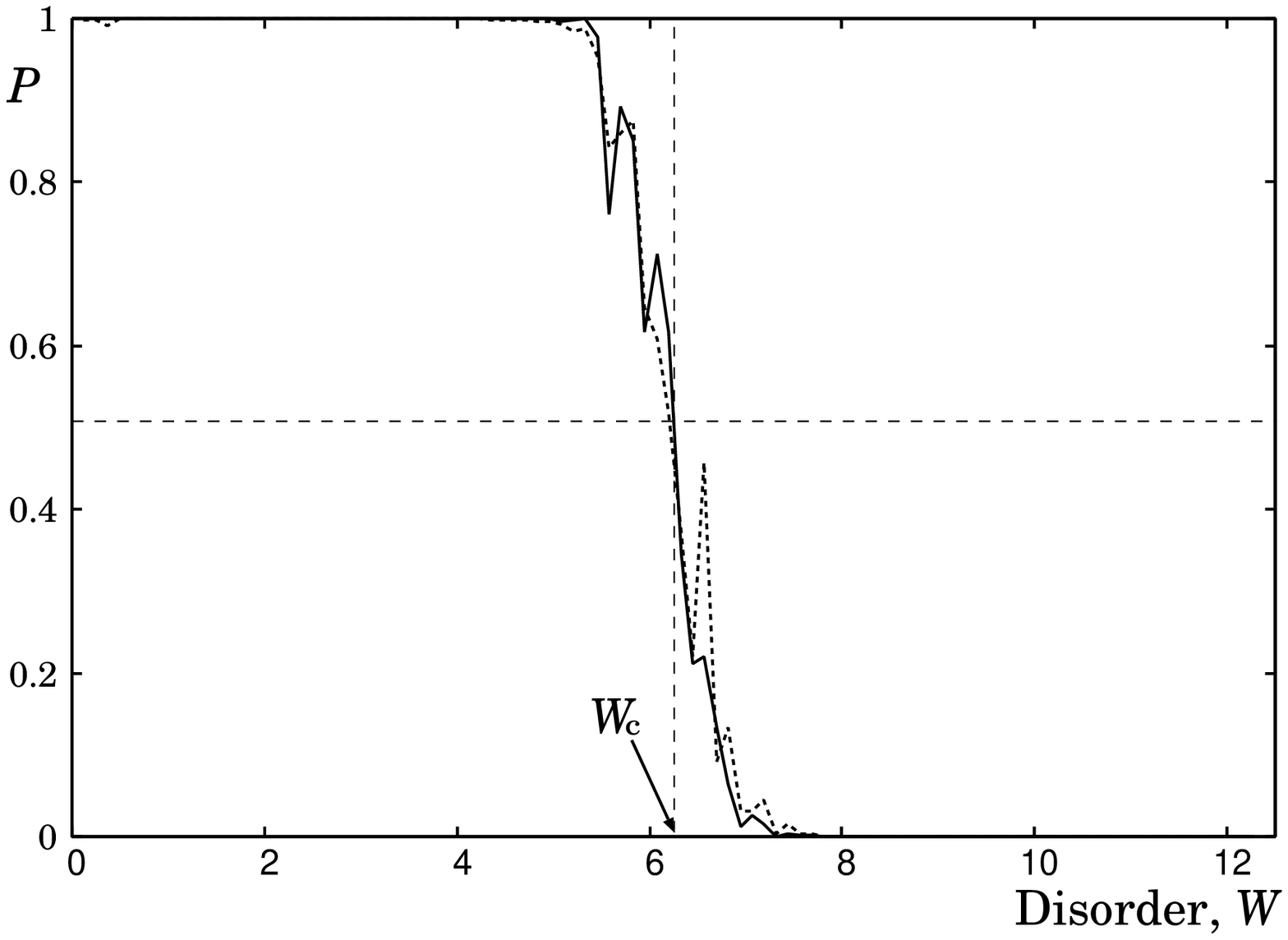}
      \hspace{1.6cm} (a)
     \end{center}
 \end{minipage}
         \begin{minipage}{0.45\hsize}
  \begin{center}
  \includegraphics[width=70mm]{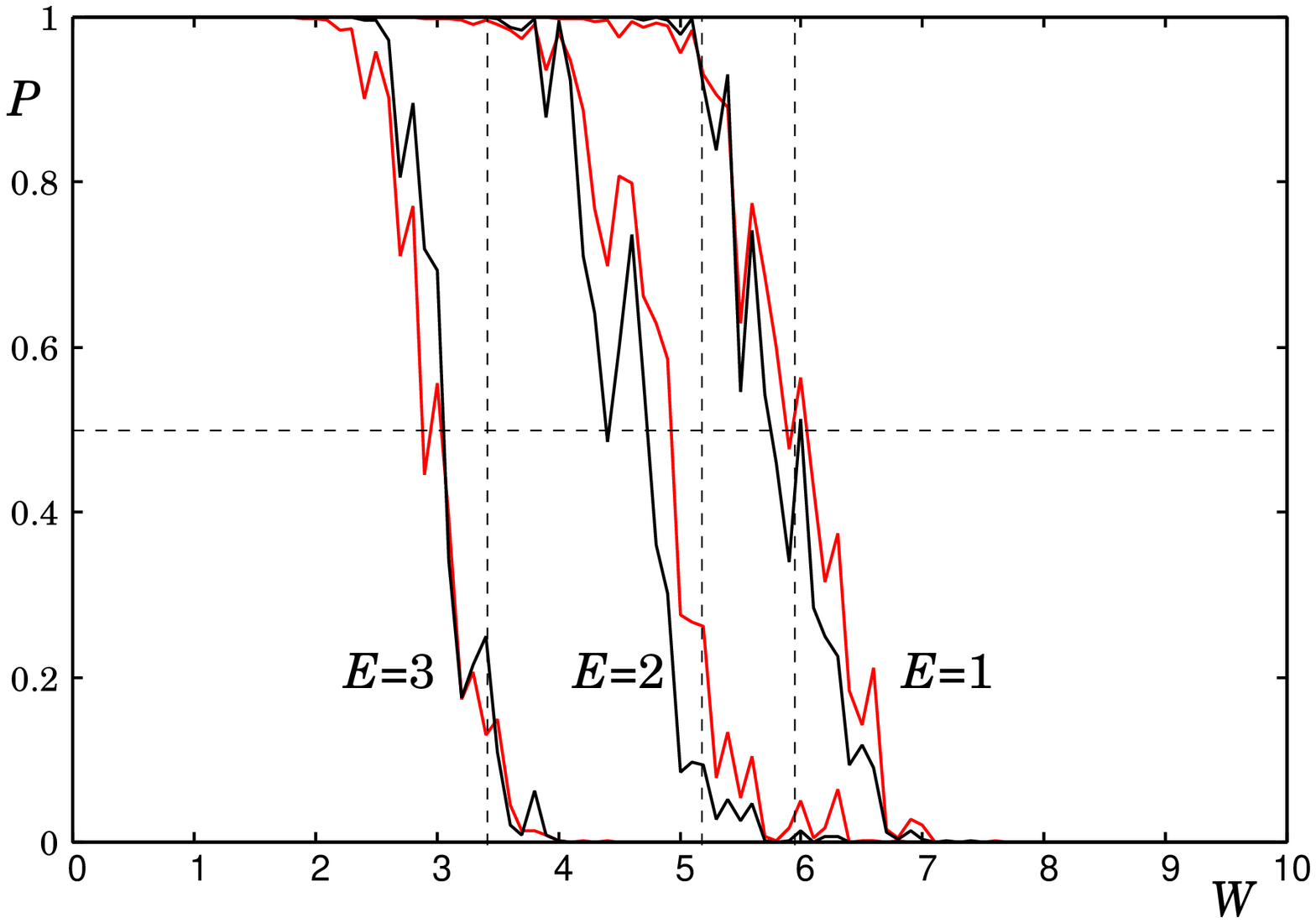}
      \hspace{1.6cm} (b)
     \end{center}
 \end{minipage}
  \end{tabular}
  \caption{(Color online) Probability of eigenfunction to be judged delocalized as
a function of disorder $W$.  Averages over 5 samples are taken.
(a) Band Center $E=0$. Critical disorder $W_c^{\rm SU2}\approx 6.20$, as well as 50\% probability,
is indicated as the dashed lines. The dotted line is for two-weight-layer network
[Fig.~\ref{fig:networkArchitectures}(a)], while the solid one is for four-weight-layer network
[Fig.~\ref{fig:networkArchitectures}(b)].
(b) For $E=1.0, 2.0$, and $3.0$.
The red line is for two-weight-layer network, while the black line is for four-weight-layer network.
The values of $W_c^{\rm SU2}$ for $E=1.0, 2.0$, and $3.0$ estimated via the finite size scaling of the localization length\cite{Asada04}
are 5.953, 5.165, and 3.394, respectively, which are indicated by the vertical dashed lines.
}
\label{fig:averagedProbability}
\end{center}
\end{figure}

We then test 5 sets of ensemble, each consisting of 100 eigenstates,
and let the machine judge whether the states are localized or not.
The resulting probability for eigenfunction to be delocalized, $P$, is shown in
Fig.~\ref{fig:averagedProbability}(a).

We then apply the results of the learning around $E=0$ 
to judge whether the states around $E=1.0, 2.0,$ and $3.0$ are delocalized.
Results are shown in Fig.~\ref{fig:averagedProbability}(b), in which we observe that,
with increasing $E$, that is, as we move from band center to band edge,
 the electron begins to be localized with a smaller strength of the disorder $W$,
 qualitatively consistent with the finite size scaling analysis.\cite{Asada04}
 There seems to be, however, a systematic deviation of the 50\% criterion of localization-delocalization
transition and the actual critical point with increasing $E$.  This may be due to the appearance of
 bound states near the band edge, which is absent in the machine learning around $E=0$.
We have further applied the results of SU(2) model machine learning for the Ando model \cite{Ando89},
and verified that
once the machine learns the eigenfunction features in certain systems, it can be applied to other systems
belonging to the same class of quantum phase transition (see Supplemental material for detail \cite{supplementalTomoki16}).

{\it Distinguishing Topological Edge States from Non-topological Ones}--
We next study the topological Chern insulator to nontopological Anderson insulator transition\cite{Dahlhaus11,Liu16,Chang16}.
We use a spinless
two-orbital tight-binding model on a square lattice,
which consists of $s$-orbital and $p\equiv p_x+ip_y$ orbital,~\cite{Qi08}
\begin{align}
H = & \sum_{{\bm x}} \left(
(\epsilon_s + v_s({\bm x})) c^{\dagger}_{{\bm x},s} c_{{\bm x},s}
+ (\epsilon_p + v_p({\bm x})) c^{\dagger}_{{\bm x},p} c_{{\bm x},p}\right)   \nonumber \\
 +& \sum_{{\bm x}}\big(-\sum_{\mu=x,y} (
t_s c^{\dagger}_{{\bm x} + {\bm e}_{\mu},s} c_{{\bm x},s}
- t_p c^{\dagger}_{{\bm x} + {\bm e}_{\mu},p} c_{{\bm x},p})  \nonumber \\
& +  t_{sp}
(c^{\dagger}_{{\bm x}+{\bm e}_x,p}
- c^{\dagger}_{{\bm x} - {\bm e}_x,p})  \!\ c_{{\bm x},s} 
-  it_{sp}
(c^{\dagger}_{{\bm x}+{\bm e}_y,p}
- c^{\dagger}_{{\bm x} - {\bm e}_y,p})  \!\ c_{{\bm x},s}
+{\rm h.c.}\big) \,,\nonumber 
\label{tb1}
\end{align}
where $\epsilon_s$,  $v_s({\bm x})$, $\epsilon_p$, and $v_p({\bm x})$ 
denote atomic energy and disorder potential for the $s$- and $p$-orbitals, respectively.
Both $v_s({\bm x})$ and $v_p({\bm x})$ 
are uniformly distributed within $[-W/2,W/2]$ with
identical and independent probability distribution. $t_s$, $t_p$, and $t_{sp}$ are
 transfer integrals between neighboring $s$-orbitals, $p$-orbitals, and that between
$s$- and $p$-orbitals, respectively.

In the absence of disorder, the system is a Chern insulator when the band inversion
condition is satisfied: $0<|\epsilon_s-\epsilon_p|<4(t_{s} + t_{p})$.
We set $\epsilon_s-\epsilon_p=-2(t_s+t_p)$, $\epsilon_s=-\epsilon_p<0$, and $t_s=t_p>0$
so that this condition is satisfied, and set $t_{sp}=4 t_s/3$.
The energy unit is set to $4t_s$.
A bulk band gap appears in $|E|<E_g=0.5$ where chiral edge states exist.

\begin{figure}[htbp]
  \begin{center}
    \begin{tabular}{c}
    
     \begin{minipage}{0.23\hsize}
  \begin{center}
   \includegraphics[width=28mm,angle=270]{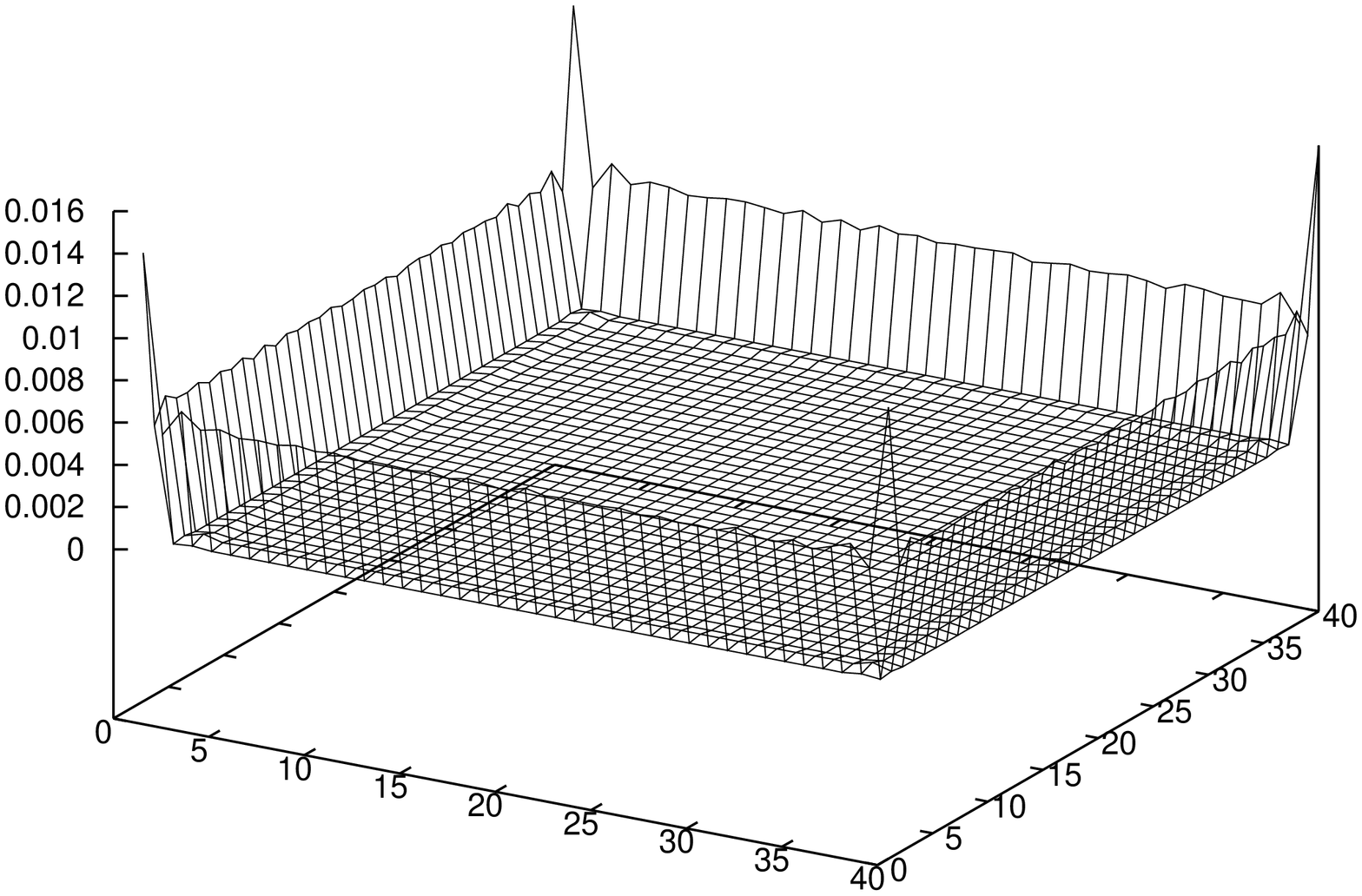}
   \hspace{1.6cm} (a)
     \end{center}
 \end{minipage}
 
 \begin{minipage}{0.23\hsize}
  \begin{center}
   \includegraphics[width=28mm,angle=270]{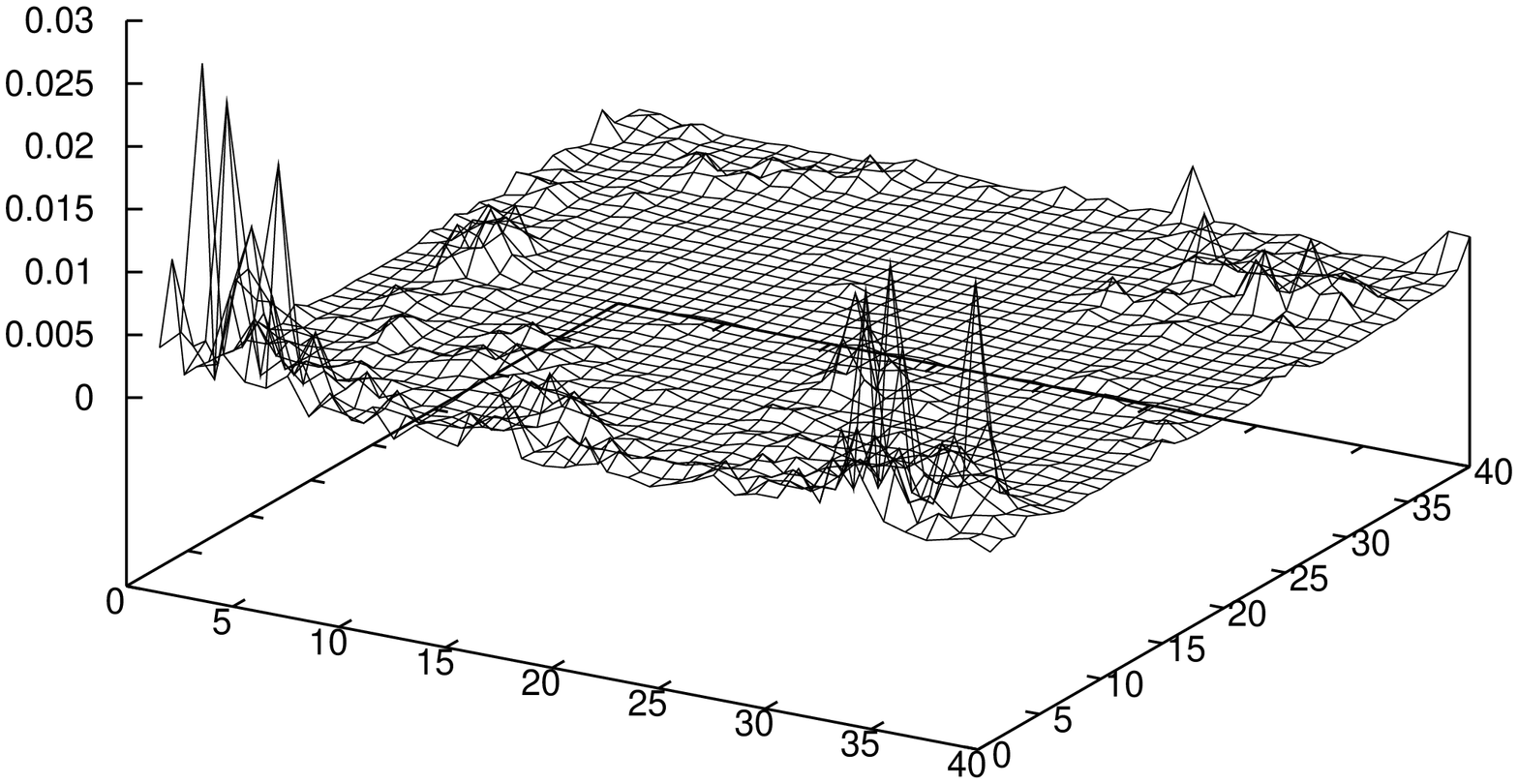}
      \hspace{1.6cm} (b)
     \end{center}
 \end{minipage}

 \begin{minipage}{0.23\hsize}
  \begin{center}
   \includegraphics[width=28mm,angle=270]{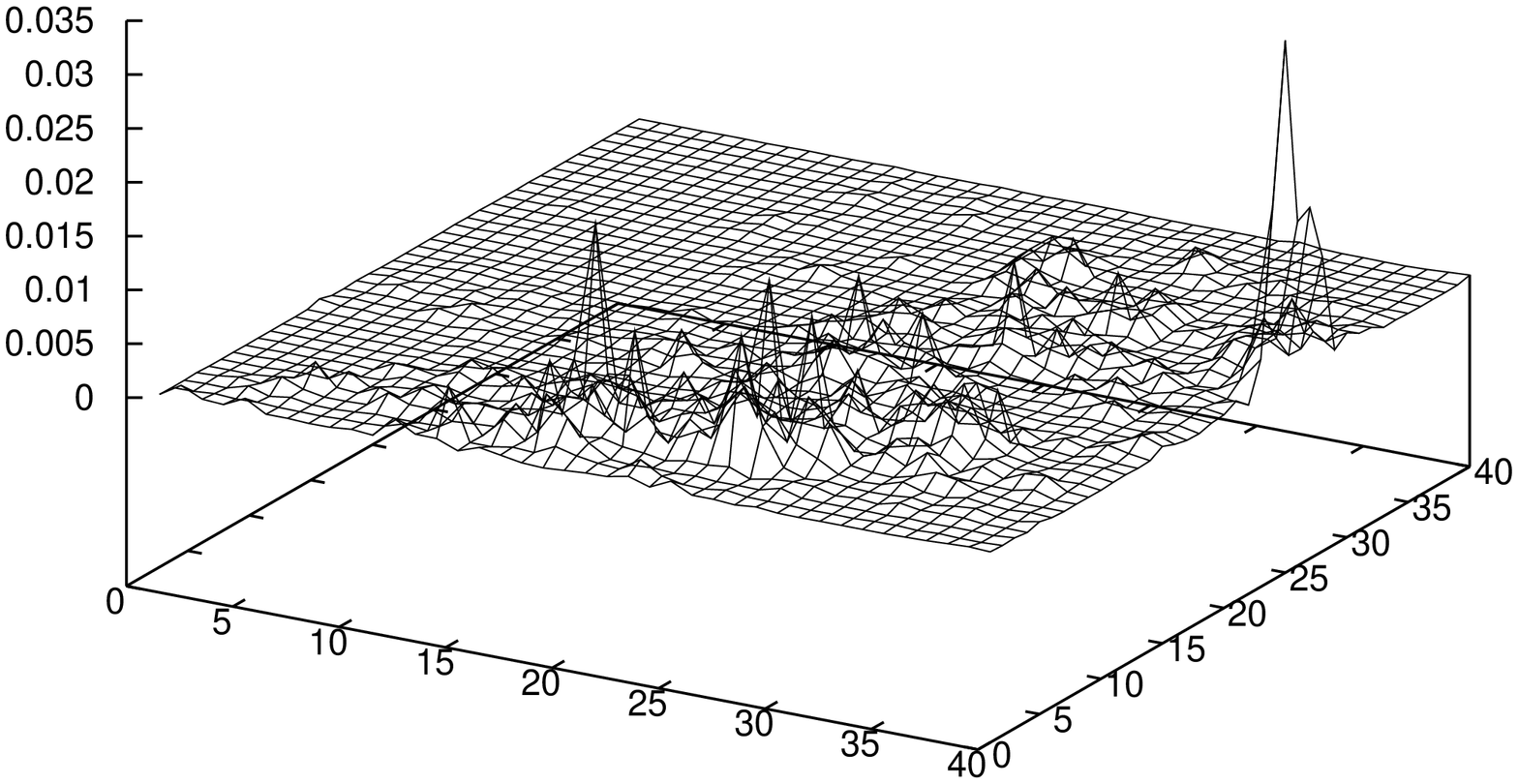}
   \hspace{1.6cm} (c)
  \end{center}
 \end{minipage}

 \begin{minipage}{0.23\hsize}
  \begin{center}
   \includegraphics[width=28mm,angle=270]{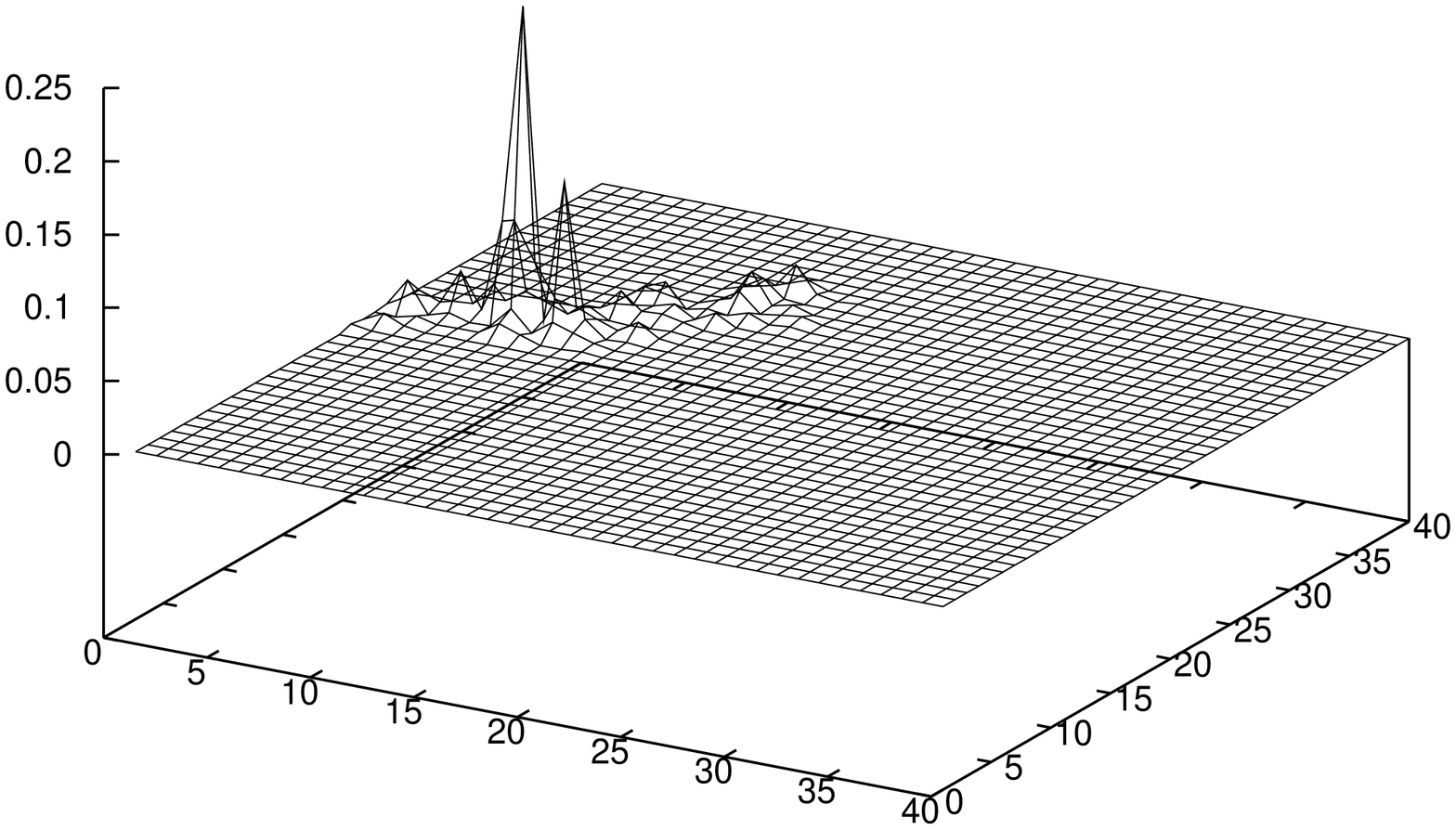}
   \hspace{1.6cm} (d)
  \end{center}
 \end{minipage}
 
  \end{tabular}
 \caption{Eigenfunction modulus squared $|\psi(x,y)|^2$ for
(a) $W=0.064\approx W_c^{\rm CI}/50$, (b) $W=2.88\approx 0.9W_c^{\rm CI}$, (c) $W=3.52\approx 1.1 W_c^{\rm CI}$, and (d) $W=6.4\approx 2W_c^{\rm CI}$.}
\label{fig:TIEigenfunctions}
\end{center}
\end{figure}

For $E=0$, the system remains as a Chern insulator for $W<W_c^{\rm CI}\approx 3.2$\cite{Liu16},
while it is an Anderson insulator for $W>W_c^{\rm CI}$.
(Unfortunately, the estimate of $W_c^{\rm CI}$ is less precise than the SU(2) model.)
We impose fixed boundary conditions in the $x$- and $y$-directions,
so that the edge states appear if the system is a topological insulator.

We diagonalize square systems of $40\times 40$ sites, and
from the resulting 3200 eigenstates,
we pick up the 1600th eigenstate.
Examples of the eigenfunctions in topological Chern [Figs.~\ref{fig:TIEigenfunctions}(a) and \ref{fig:TIEigenfunctions}(b)] and nontopological Anderson insulators [Figs.~\ref{fig:TIEigenfunctions}(c) and \ref{fig:TIEigenfunctions}(d)]
are shown in Fig.~\ref{fig:TIEigenfunctions}.
As shown in Fig.~\ref{fig:TIEigenfunctions}, it is difficult to judge whether
the state is an edge state or not when $W$ is close to $W_c^{\rm CI}$: see, for example,
$W=0.9W_c^{\rm CI}$ [Fig.~\ref{fig:TIEigenfunctions}(b), Chern insulator phase] and $W=1.1W_c^{\rm CI}$ 
[Fig.~\ref{fig:TIEigenfunctions}(c), Anderson insulator phase].
In fact, learning 1000 samples for each phase gives 93\% validation accuracy for four-weight-layer network compared
with 98\% or more as in the SU(2) model.
The difficulty may be due to the fixed boundary condition where shifting the locus of the maximum of the eigenfunction
amplitude is not allowed.
Another reason for difficulty is that the bulk of the systems are localized in both topological and
nontopological regions.
To overcome these difficulties, we increased the number of samples: 27000 samples belonging to the topological phase, and  27000 to the nontopological phase.
We have also increased the number of hidden units to be 32 for the first convolution layer (``conv1'' in Fig.~\ref{fig:networkArchitectures}), 128 for the second (``conv2''), and 512 for the hidden dense connection layer (``ip1''). 

\begin{figure}
  \begin{center}
    \begin{tabular}{c}
         \begin{minipage}{0.4\hsize}
  \begin{center}
   \includegraphics[width=60mm]{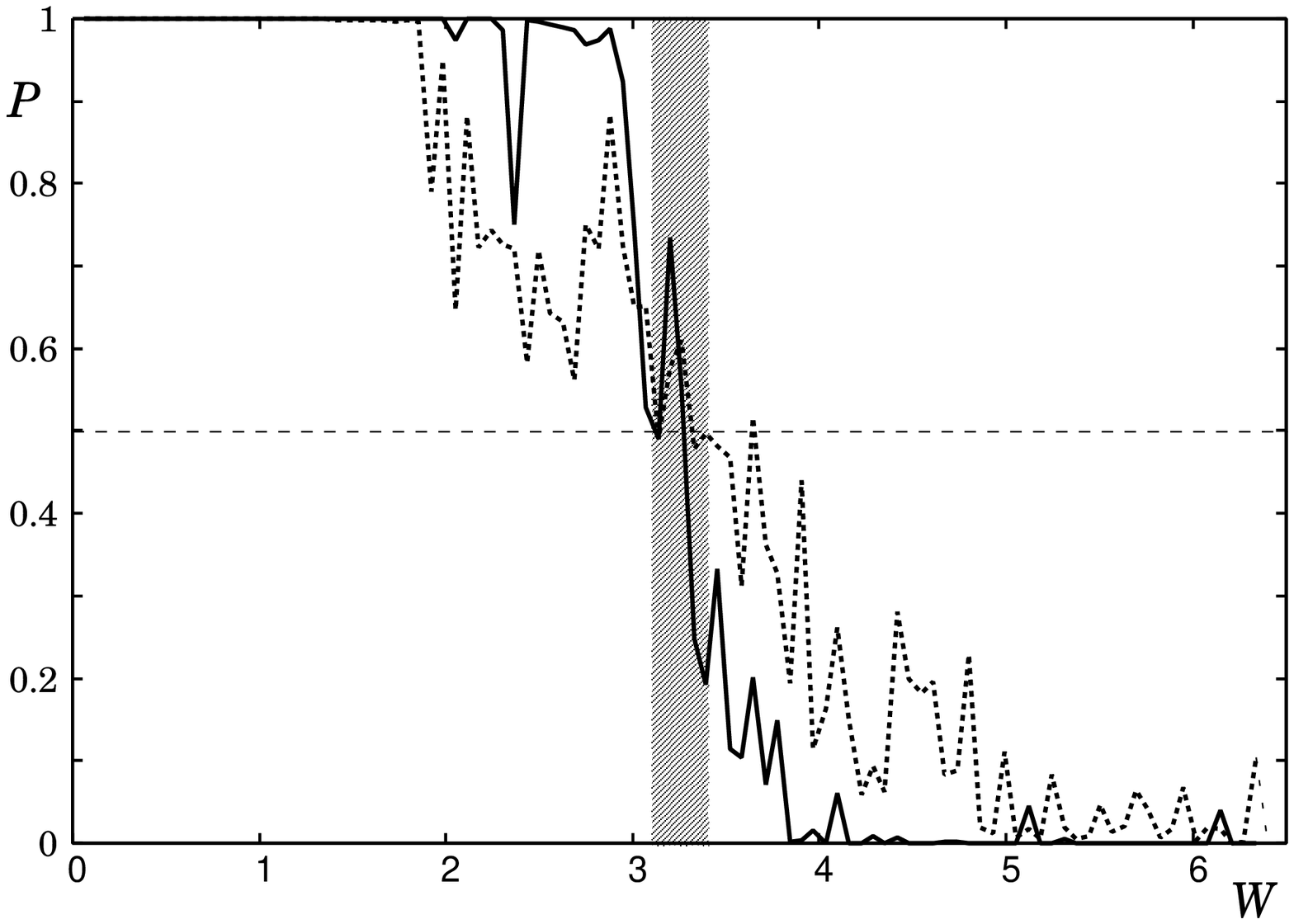}
      \hspace{1.6cm} (a)
     \end{center}
 \end{minipage}
         \begin{minipage}{0.4\hsize}
  \begin{center}
  \includegraphics[width=60mm]{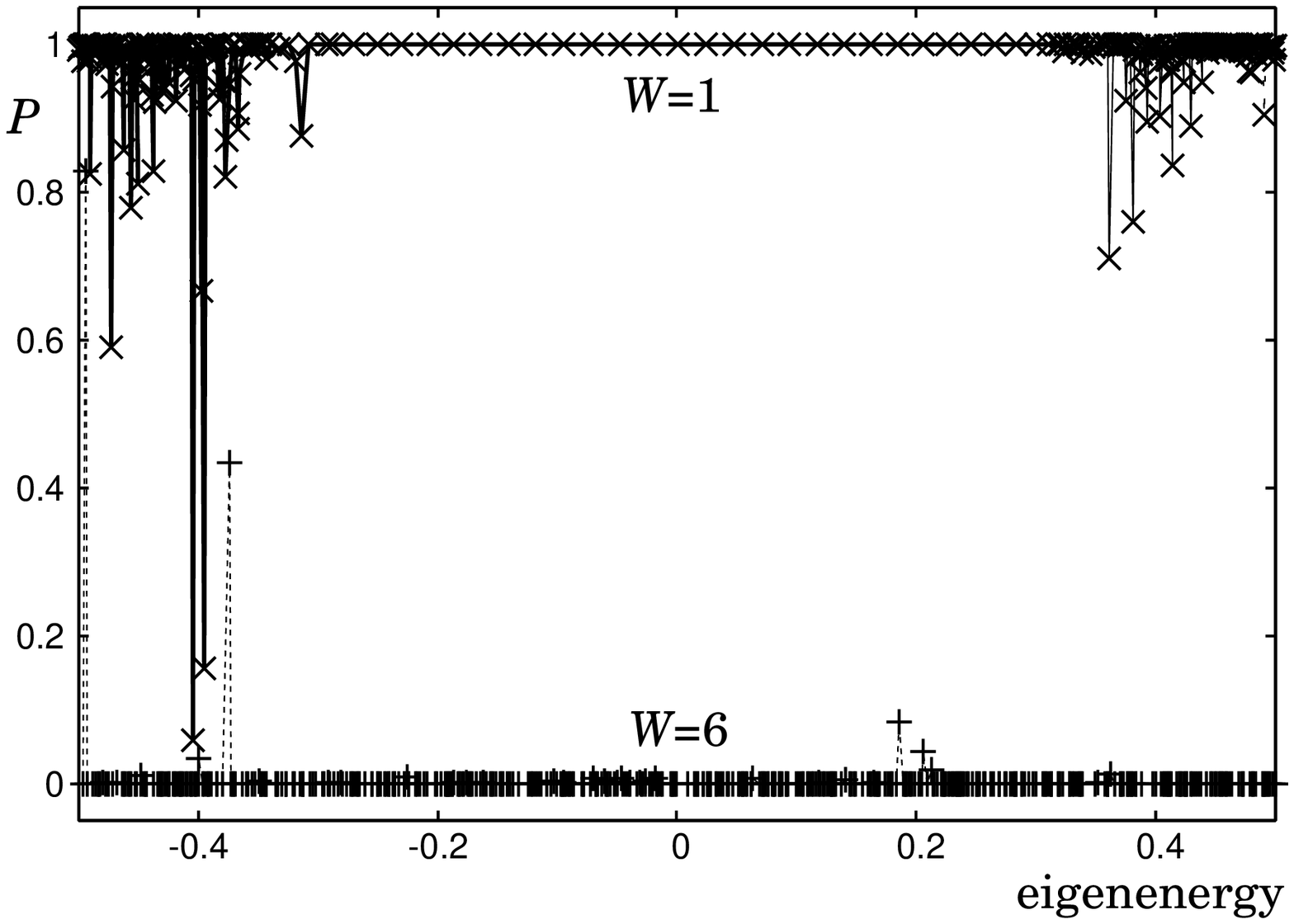}
      \hspace{1.6cm} (b)
     \end{center}
 \end{minipage}
  \end{tabular}
 \caption{(a) Probability of eigenfunction around $E=0$ to be judged topological
edge states as
a function of disorder $W$.  Averages over 5 samples are taken.
50\% probability is indicated as the horizontal dashed line.
Since the critical disorder is less accurate, $W_c^{\rm CI}=3.25\pm 0.1$
is shown as a shaded region.
The dotted line is for a two-weight-layer network, while
the solid one is for a four-weight-layer one.
(b) Same quantity but as a function of eigenenergy $E$ inside 
the bulk band gap region $|E|<E_g=0.5$.
Results for $W=1<W_c^{\rm CI}$ ($\times$, solid line) and $W=6>W_c^{\rm CI}$($+$, dotted line)
are shown.}
\label{fig:averagedProbabilityCI}
\end{center}
\end{figure}

In Fig.~\ref{fig:averagedProbabilityCI}(a), we plot the probability of the eigenfunction
to be judged topological.  A new ensemble of eigenfunctions with different random number
sequences has been prepared to test this method.
As in the case of delocalization-localization transition, the probability fluctuates
near the critical point and vanishes in the nontopological region.
The validation accuracy is 90\% for the case of two layers of network (dotted line),
and 97\% for four layers of network (solid line), which
demonstrates clearly that a deeper network exhibits better performance.

We next apply the result of the deep learning around $E=0$ to judge the states in the bulk band gap region at zero disorder, $|E|<E_g=0.5$.
We diagonalize a system for $W=1<W_c^{\rm CI}$ and $W=6>W_c^{\rm CI}$, take
all the eigenstates with  $|E|<E_g$, and let the machine judge them.
Figure \ref{fig:averagedProbabilityCI}(b) shows that topological edge states other than $E=0$
are also well distinguished from nontopological ones based on the learning around $E=0$.

{\it Concluding Remarks}--
In this paper, we focused on 2D random electron systems.
We have demonstrated the validity of deep learning for distinguishing various random electron
states in quantum phase transitions.
For strong enough and weak enough randomness, the precision of judgement is 0.99999$\cdots$,
while in the critical regime, the judgement becomes less accurate.
This region is related to the critical region where the characteristic length scale $\xi$ is comparable to or
longer than the system size $L$. 
That is, the probability $P$ for the eigenfunction to be judged delocalized/topological obeys
the scaling law, $P(W, L)=f[(W-W_c)L^{1/\nu}]$, although determining the exponent $\nu$
is beyond the scope of this Letter.
Since all we need to calculate are eigenfunctions with relatively small systems,
the method will work for systems where the transfer matrix method is not applicable
(localization problems on random\cite{Avishai92,Berkovits96,Kaneko99,Ujfalusi14}
and fractal lattices\cite{Asada06}, for example).

We have used the known values of critical disorder to teach the machine.
After learning the feature of eigenfunctions near the band center, the machine
could capture localized/delocalized and topological/nontopological features away from the band center.
We have also verified that the results of the SU(2) model learning can be applied to the Ando model.\cite{supplementalTomoki16}

In the cases of Anderson transition near the band edge in the SU(2) model [Fig.~\ref{fig:averagedProbability}(b)]
and that at the band center in the Ando model,
the machine tends to predict
the transition for a slightly smaller disorder than the estimate of finite size scaling analyses \cite{Ando89,Fastenrath91}.
We have extracted the features in the middle layers to explain this tendency,\cite{supplementalTomoki16}
but could not clarify how the machine judges phases.
The details of judgement should be clarified in the future.

We have focused on the amplitude of eigenfunction in 2D.
In higher dimensions, the same algorithm will be applicable
via dimensional reduction: integration of $|\psi^2|$ over certain directions, reducing the image to two dimensions.
The dimensional reduction will also work for disordered 3D strong and weak topological
insulators.\cite{Kobayashi13}
Other interesting quantities for machine learning are phase and spin texture of eigenfunctions in random electron systems.
Classical waves (photon, phonon) in random media\cite{shengbook,aegerter06,faez09}
as well as disordered magnon\cite{Xu16}
are also worth machine learning.

\begin{acknowledgments}
The authors would like to thank Keith Slevin, Koji Kobayashi, and Ken-Ichiro Imura for useful discussions.
This work was partly supported by JSPS KAKENHI Grant No. JP15H03700.
\end{acknowledgments}

\bibliography{tomokiJPSJ2016}

\begin{thebibliography}{10}

\bibitem{Anderson58}
P.~W. Anderson: Phys. Rev. {\bfseries 109} (1958) 1492.

\bibitem{Schnyder08}
A.~P. Schnyder, S.~Ryu, A.~Furusaki, and A.~W.~W. Ludwig: Phys. Rev. B
  {\bfseries 78} (2008) 195125.

\bibitem{Kitaev09}
A.~Kitaev: AIP Conference Proceedings {\bfseries 1134} (2009) 22.

\bibitem{Hasan10}
M.~Z. Hasan and C.~L. Kane: Reviews of Modern Physics {\bfseries 82} (2010)
  3045.

\bibitem{Qi11}
X.-L. Qi and S.-C. Zhang: Rev. Mod. Phys. {\bfseries 83} (2011) 1057.

\bibitem{Zirnbauer96}
M.~R. Zirnbauer: Journal of Mathematical Physics {\bfseries 37} (1996) 4986.

\bibitem{altland97}
A.~Altland and M.~R. Zirnbauer: Phys. Rev. B {\bfseries 55} (1997) 1142.

\bibitem{Sbierski14a}
B.~Sbierski and P.~W. Brouwer: Physical Review B {\bfseries 89} (2014) 155311.

\bibitem{Katsura16}
H.~Katsura and T.~Koma: J. Math. Phys. {\bfseries 57} (2016) 021903.

\bibitem{Katsura16a}
H.~Katsura and T.~Koma: arXiv:1611.01928  (2016).

\bibitem{alberto10}
A.~Rodriguez, L.~J. Vasquez, K.~Slevin, and R.~A. Romer: Phys. Rev. Lett.
  {\bfseries 105} (2010) 046403.

\bibitem{alberto11}
A.~Rodriguez, L.~J. Vasquez, K.~Slevin, and R.~A. Romer: Phys. Rev. B
  {\bfseries 84} (2011) 134209.

\bibitem{Slevin14}
K.~Slevin and T.~Ohtsuki: New Journal of Physics {\bfseries 16} (2014) 015012.

\bibitem{Ujfalusi15}
L.~Ujfalusi and I.~Varga: Physical Review B {\bfseries 91} (2015) 184206.

\bibitem{Obuchi14}
T.~Obuchi, H.~Koma, and M.~Yasuda: Journal of the Physical Society of Japan
  {\bfseries 85} (2016) 114803.

\bibitem{LeCun15}
Y.~LeCun, Y.~Bengio, and G.~Hinton: Nature {\bfseries 521} (2015) 436.

\bibitem{Silver16}
D.~Silver~{\it et al.}: Nature {\bfseries 529} (2016) 484.

\bibitem{Carrasquilla16}
J.~Carrasquilla and R.~G. Melko: arXiv:1605.01735  (2016).

\bibitem{Tanaka16}
A.~Tanaka and A.~Tomiya: arXiv:1609.09087  (2016).

\bibitem{Carleo16}
G.~Carleo and M.~Troyer: arXiv:1606.02318  (2016).

\bibitem{Broecker16}
P.~Broecker, J.~Carrasquilla, R.~G. Melko, and S.~Trebst: arXiv:1608.07848
  (2016).

\bibitem{Chng16}
K.~Ch'ng, J.~Carrasquilla, R.~G. Mello, and E.~Khatami: arXiv:1609.02552
  (2016).

\bibitem{Li16}
L.~Li, T.~E. Baker, S.~R. White, and K.~Burke: arXiv:1609.03705  (2016).

\bibitem{Nieuwenburg16}
E.~P. van Nieuwenburg, Y.-H. Liu, and S.~D. Huber: arXiv:1610.02048  (2016).

\bibitem{Huang16}
L.~Huang and L.~Wang: arXiv:1610.02746  (2016).

\bibitem{Asada02}
Y.~Asada, K.~Slevin, and T.~Ohtsuki: Phys. Rev. Lett. {\bfseries 89} (2002)
  256601.

\bibitem{Asada04}
Y.~Asada, K.~Slevin, and T.~Ohtsuki: Physical Review B {\bfseries 70} (2004)
  035115.

\bibitem{lecun1998gradient}
Y.~LeCun, L.~Bottou, Y.~Bengio, and P.~Haffner: Proceedings of the IEEE
  {\bfseries 86} (1998) 2278.

\bibitem{jia2014caffe}
Y.~Jia, E.~Shelhamer, J.~Donahue, S.~Karayev, J.~Long, R.~Girshick,
  S.~Guadarrama, and T.~Darrell: arXiv:1408.5093  (2014).

\bibitem{glorot2010understanding}
X.~Glorot and Y.~Bengio: Aistats {\bfseries 9} (2010) 249.

\bibitem{tieleman2012lecture}
T.~Tieleman and G.~Hinton: COURSERA: Neural Networks for Machine Learning
  {\bfseries 4} (2012) Lecture 6.5.

\bibitem{Ando89}
T.~Ando: Phys. Rev. B {\bfseries 40} (1989) 5325.

\bibitem{supplementalTomoki16}
Results for Ando model and intermediate feature maps are provided online as
  Supplemental material.

\bibitem{Dahlhaus11}
J.~P. Dahlhaus, J.~M. Edge, J.~Tworzyd\l{}o, and C.~W.~J. Beenakker: Phys. Rev.
  B {\bfseries 84} (2011) 115133.

\bibitem{Liu16}
S.~Liu, T.~Ohtsuki, and R.~Shindou: Phys. Rev. Lett. {\bfseries 116} (2016)
  066401.

\bibitem{Chang16}
C.-Z. Chang, W.~Zhao, J.~Li, J.~K. Jain, C.~Liu, J.~S. Moodera, and M.~H.~W.
  Chan: Phys. Rev. Lett. {\bfseries 117} (2016) 126802.

\bibitem{Qi08}
X.-L. Qi, T.~L. Hughes, and S.-C. Zhang: Phyical Review B {\bfseries 78} (2008)
  195424.

\bibitem{Avishai92}
Y.~Avishai and J.~M. Luck: Phys. Rev. B {\bfseries 45} (1992) 1074.

\bibitem{Berkovits96}
R.~Berkovits and Y.~Avishai: Phys. Rev. B {\bfseries 53} (1996) R16125.

\bibitem{Kaneko99}
A.~Kaneko and T.~Ohtsuki: Journal of the Physical Society of Japan {\bfseries
  68} (1999) 1488.

\bibitem{Ujfalusi14}
L.~Ujfalusi and I.~Varga: Phys. Rev. B {\bfseries 90} (2014) 174203.

\bibitem{Asada06}
Y.~Asada, K.~Slevin, and T.~Ohtsuki: Phys. Rev. B {\bfseries 73} (2006) 041102.

\bibitem{Fastenrath91}
U.~Fastenrath, G.~Adams, R.~Bundschuh, T.~Hermes, B.~Raab, I.~Schlosser,
  T.~Wehner, and T.~Wichmann: Physica A {\bfseries 172} (1991) 302.

\bibitem{Kobayashi13}
K.~Kobayashi, T.~Ohtsuki, and K.-I. Imura: Physical Review Letters {\bfseries
  110} (2013) 236803.

\bibitem{shengbook}
P.~Sheng: {\em Introduction to wave scattering, localization, and mesoscopic
  phenomena} (Academic Press, San Diego, 1995).

\bibitem{aegerter06}
C.~M. Aegerter, M.~St\"orzer, and G.~Maret: Europhys. Lett. {\bfseries 75}
  (2006) 562.

\bibitem{faez09}
S.~Faez, A.~Strybulevych, J.~H. Page, A.~Lagendijk, and B.~A. van Tiggelen:
  Phys. Rev. Lett. {\bfseries 103} (2009) 155703.

\bibitem{Xu16}
B.~Xu, T.~Ohtsuki, and R.~Shindou: arXiv:1606.02839  (2016).

\bibitem{MacKinnon83}
A.~MacKinnon and B.~Kramer: ZEITSCHRIFT FUR PHYSIK B-CONDENSED MATTER
  {\bfseries 53} (1983) 1.

\end{thebibliography}

\newpage

\begin{center}
{\large Supplemental Material for ``Deep Learning the Quantum Phase Transitions in Random Two-Dimensional Electron Systems''}
\end{center}

\begin{figure}[htb]
  \begin{center}
  \includegraphics[width=70mm]{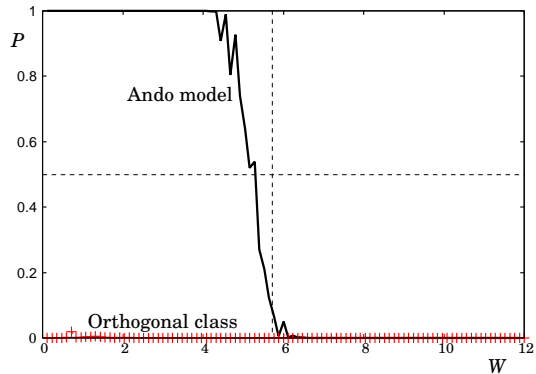}
\caption{Probability of eigenfunction of the Ando model to be judged delocalized as
a function of disorder $W$ based on SU(2) machine learning.  Averages over 5 samples are taken.
$W_c^{\rm Ando} \approx 5.75$ is indicated as a vertical dashed line.
Results for orthogonal class (red $+$) are also shown.}
\label{fig:AndoModel}
     \end{center}
\end{figure}

\noindent
{\it Application of SU(2) learning to Ando model}--
 To further confirm the validity of machine learning demonstrated in the main text, we have applied the results of the SU(2) model learning 
 to another model that describes the constant strength of spin-orbit coupling, i.e., Ando model\cite{Ando89}
 (Fig.~\ref{fig:AndoModel}).
 In this model, $\alpha$ in Eq.(2) in the main text is set to 0, $\gamma$ is $0$ for $x$-direction transfer and $\pi/2$
 for $y$-direction.  We set the strength of the spin-orbit coupling $\beta = \pi/6$
 to compare with the previous results \cite{Ando89, Fastenrath91},
  $W_c^{\rm Ando}\approx 5.75$.  The solid line shows that the features of localization-delocalization transition
  learned from a model (SU(2)) can be applied to a different model (Ando), though the probability of delocalization
  starts to decrease with increasing $W$ slightly earlier than expected.
  This might be due to the corrections to scaling, which is present in Ando model but
   negligible in SU(2) model.
  
  We have also set $\beta=0$ (no spin-orbit coupling, i.e., the Anderson model, which belongs to the orthogonal class), where all the states are expected to be localized,
  which is actually the case of machine judgement (red $+$). 
  The machine judgement is, however, too good in small disorder region $W<5$
  where the localization length becomes greater than 100 lattice cites \cite{MacKinnon83},

  larger than the system size 40 .
  This may be due to the standing wave like structure of eigenfunctions in this region, where the peak values
  are fluctuating due to disorder, from which the eigenfunctions might have been judged to be localized.

\begin{figure}[htb]
  \begin{center}
  \includegraphics[width=0.92\textwidth]{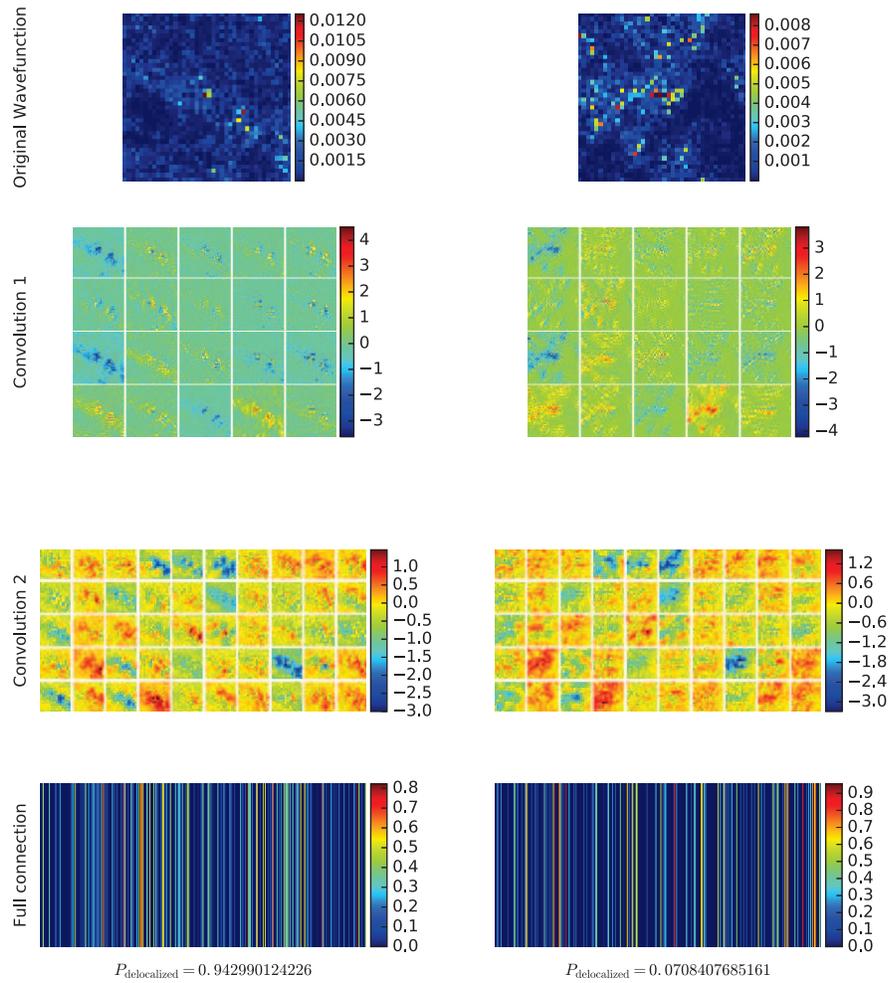}
\caption{Features in the intermediate layers for states in the SU(2) model.
The top two panels show the modulus squared of eigenfunctions.
Results of the 1st and the 2nd convolutions are shown in
the 2nd and the 3rd rows, followed by the features of
the final full connection.
The left column shows how a state in a delocalized region
is judged to be delocalized with probability 0.9429..., while the right
one shows how a state in the localized region is judged to be
localized with delocalization probability 0.0708...}
\label{fig:locDelocFeature}
     \end{center}
\end{figure}

\begin{figure}[htb]
  \begin{center}
  \includegraphics[width=0.92\textwidth]{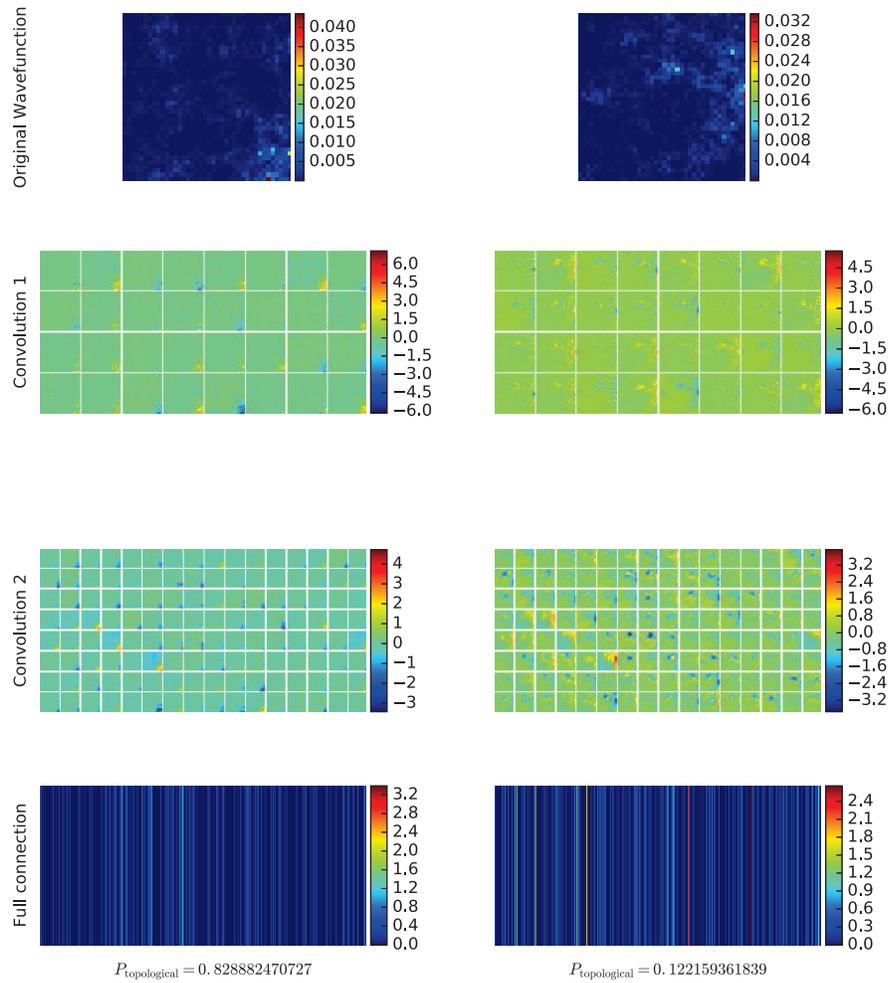}
\caption{Features in the intermediate layers for states in a Chern insulator.
The top, the 2nd, the 3rd and the 4th rows mean the same as in the previous figure.
The left column shows how a state in a topological insulator phase
is judged to be a topological edge state with probability 0.8288...,
while the right one shows how a state in the Anderson insulator phase is judged to be
a topological edge state with probability 0.1221...}
\label{fig:tiNonTiFeature}
     \end{center}
\end{figure}

\newpage
{\it Features in the intermediate layers}--
Here we show examples of features in the intermediate layers for
localization-delocalization transition (Fig.~\ref{fig:locDelocFeature})
and topological-nontopological transition  (Fig.~\ref{fig:tiNonTiFeature}).

\end{document}